\documentclass[11pt]{article}
\usepackage{geometry}                
\usepackage{graphicx}
\usepackage{graphics}
\usepackage{amssymb}
\usepackage{amsmath}
\usepackage{epstopdf}
\usepackage{rotating}
\usepackage{subfigure} 
\usepackage{multirow}
\usepackage[round,authoryear]{natbib}
\usepackage{color}
\let\cite\citep

\title{\textbf{The Mittag-Leffler Fitting of the Phillips Curve}}
\author{Tomas Skovranek}
\date{}                                           

\begin{document}
\maketitle

\begin{center}
\textit{Institute of Control and Informatization of Production Processes\\
        BERG Faculty, Technical University of Kosice\\
        Nemcovej 3, 042 00 Kosice, Slovak Republic\\
        phone number: +421556025143 \\
        e-mail: tomas.skovranek@tuke.sk	
}
\end{center}

\vspace{0.5cm}
                       
\begin{abstract}

In this paper, a~mathematical model based on the one-parameter Mittag-Leffler function is proposed to be used for the first time to describe the relation between unemployment rate and inflation rate, also known as the Phillips curve. The Phillips curve is in the literature often represented by an exponential-like shape. On the other hand, Phillips in his fundamental paper used a~power function in the model definition. Considering that the ordinary as well as generalised Mittag-Leffler function behaves between a~purely exponential function and a~power function it is natural to implement it in the definition of the model used to describe the relation between the data representing the Phillips curve. For the modelling purposes the data of two different European economies, France and Switzerland, were used and an ``out-of-sample'' forecast was done to compare the performance of the Mittag-Leffler model to the performance of the power-type and exponential-type model. The results demonstrate that the ability of the Mittag-Leffler function to fit data that manifest signs of stretched exponentials, oscillations or even damped oscillations can be of use when describing economic relations and phenomenons, such as the Phillips curve. 

\end{abstract}


\section{Introduction}
It is ``because of'' or ``thanks to'' Paul Anthony Samuelson and Robert Merton Solow \cite{Samuelson1960}, that the economists all around the world call the negative correlation between the rate of wage change (or the price inflation rate) and unemployment rate the Phillips curve (PC). It is lesser-known, that the idea occurred more than 30 years before publishing  the famous paper of Alban William Housego Phillips \cite{Phillips1958}, in the work by Irving Fisher \cite{Fisher1926}. And Fisher was not the only one who would deserve such an important ``discovery'' be named after him. Arthur Joseph Brown in his book \cite{Brown1955}, published 3 years before Phillips paper, precisely described the inverse relation between the wage and price inflation and the rate of unemployment. Also Richard George Lipsey \cite{Lipsey1960} played an important role
by the birth, creation of the theoretical foundations and popularisation of the PC. Samuelson and Solow in their paper  \cite{Samuelson1960} for the first time mention policy implications.
In the empirical studies \cite{Brown1955, Phillips1958,Lipsey1960} for the United Kingdom, and \cite{Bowen1960, Samuelson1960, Bodkin1966} for the United States, the inverse relationship between the rate of wage change and the unemployment rate was proven. The PC was in its beginnings widely used by  the policy-makers to exploit the trade-off to reduce unemployment at a small cost of additional inflation - ``sacrifice ratio''.

Since then the PC has been studied, extended and re-formulated by many authors. One of the ``modern'' forms of the PC is represented by models in which expectations are not anchored in backward-looking behaviour but can jump in response to current and anticipated changes in policy  - the New Keynesian theory of the output-inflation tradeoff.
The model called the New Keynesian Phillips Curve (NKPC) builds, among others, on the works of Taylor \cite{Taylor1979, Taylor1980}
and Calvo \cite{Calvo1983}, where the models of staggered contracts were developed, and on the quadratic price adjustment cost model of Rotemberg and Woodford  \cite{Rotemberg1982, Rotemberg1997}, all of which have a~similar formulation as the expectations-augmented PC of Friedman and Phelps  \cite{Friedman1968, Phelps1968}.
The work of Clarida \textit{et al.} \cite{Clarida1999} illustrates the widely usage of this model in theoretical analysis of monetary policy. With the focus shifted from the unemployment rate to the output gap, Phillips' relationship has become an aggregate supply curve, but the idea remains the same.  McCallum \cite{McCallum1997} has called it ``the closest thing there is to a standard specification''.
The NKPC stayed popular also in the late 90's and at the beginning of the 21\textsuperscript{st} century as a~theory for understanding inflation dynamics.
In the works  \cite{Gali1999, Gali2001, Gali2005}
the NKPC was transformed into a hybrid version, that relates inflation to expected future inflation, lagged inflation and real marginal costs. 

When Magnus Gustaf Mittag-Leffler in his works \cite{ML1903b, ML1903a} proposed a new function $E_{\alpha}(x)$, he surely did not expect how important generalisation of the exponential function $e^{x}$ he developed. 
The ML function and its generalisations  interpolate between a purely exponential law and a~power-law-like behaviour, and they arise naturally in the solution of fractional-order integro-differential  equations, random walks, L\'{e}vy flights, the study of complex systems, and in other fields.
In numerous works the properties, generalisations and applications of the ML-type functions were studied \textit{e.g.}~ \cite{Hille1930, Dzhrbashyan1966a, Caputo1971, Blair1974, Torvik1984, Samko1993, Gorenflo1991, Gorenflo1996, Gorenflo1997d, Gorenflo2002,Gorenflo2014, Kilbas1995, Kilbas1996, Kilbas2004, Mainardi1996, Mainardi2000, Srivastava2001, Srivastava2009, West2003, Haubold2011, Garrappa2015}, and computation procedures for evaluating the ML function were developed \textit{e.g.} in \cite{Gorenflo2002, ChenML, ChenML2, PodlubnyML, PodlubnyMLfit, GarrappaMAT, MatychynML}.

The Mittag-Leffler (ML) function become of great use and importance
not only for mathematicians, but because of its special properties and huge potential by solving applied problems it found its applicability also in the fields such as psychorheology \cite{Blair1974}, electrotechnics \cite{Oliveira2011, Petras2012, Sierociuk2013}, modeling of processes such as diffusion \cite{Mainardi2018}, combustion \cite{Samuel2016}, universe expansion \cite{Zeng2015}, \textit{etc.}
The idea to use the fractional-order calculus and the ML function for modelling phenomenons from the fields of economics and econophysics was elaborated by several authors
\cite{Scalas2000, Gorenflo2001, Mainardi2000a, Mainardi2002, Cartea2007, Mendes2008, Garibaldi2010, Skovranek2012, Tejado2015, Tejado2018, Tarasov2016, Tarasov2017, Tarasova2018, Tarasov2018, Tarasov2019, Tarasov2019a, Tarasov2019b, Tarasov2019c}.

In this paper 
the one-parameter ML function
is for the first time used to model the relation between the unemployment rate and the inflation rate - the Phillips curve, and its performance is compared to the power-type model and the exponential-type model.
French and Swiss econometric data are taken for the period of time 1980 - 2017 from the portal \emph{EconStats\textsuperscript{TM}} \cite{econstats} to identify the PC of these economies.
The dataset is split into two subsets, the ``modelling'' subset is used to identify the model parameters, and a shorter ``out-of-sample'' subset serves for evaluating the forecast-performance of the models.
The performance of all three models is evaluated based on the fitting-criterion, \textit{i.e.} the sum of squared errors (SSE). The results are presented in the form of figures and tables, where the SSE of the fitting curve to the ``modelling'' subset, SSE of the fitting curve to the ``out-of-sample'' subset, and SSE of the fitting curve to the complete dataset, as well as some other quality criterions for the goodness-of-fit are listed. 

The paper is organised as follows. Section~2 gives an overview of Mittag-Leffler function and its generalisations. The original Phillips curve  as well as the and Mittag-Leffler model for fitting the Phillips curve is described in Section~3.
The numerical results and the discussion on the experiments can be found in Section~4. Finally, concluding remarks are given in Section~5.

\section{Preliminaries: Mittag-Leffler function and its generalisations}

In 1903 M. G. Mittag-Leffler \cite{ML1903b, ML1903a} introduced a new function $E_{\alpha}(x)$, a generalisation of the classical exponential function $e^x$, which is till today known as the one-parameter ML function. Using Erd\'{e}lyi's notation \cite{Erdelyi1955}, where $z$ is used instead of $x$, the function can be written as:
\begin{align}\label{eq:1term-ML}
    E_{\alpha}(z) = \sum_{k=0}^{\infty}\frac{z^{k}}
                                 {\Gamma(\alpha k+1)}, \quad
                                 \alpha \in \mathbb C,\, \operatorname{Re} (\alpha)>0,\, z \in \mathbb C,
\end{align}
where $\Gamma$ denotes the (complete) Gamma function, having the property $\Gamma(n) = (n-1)!$.
The one-parameter ML function and its properties were further investigated 
in \cite{ML1904, ML1905, Wiman1905a,Wiman1905b, Pollard1948, Agarwal1953, Humbert1953a, Humbert1953}
followed by the generalisation to a two-parameter function of the ML-type, by some authors called the Wiman's function (some give the credit to Agarwal). Following the Erd\'{e}lyi's handbook the formula has the form \cite{Erdelyi1955}:
 \begin{align}\label{eq:2term-ML}
    E_{\alpha, \beta}(z) = \sum_{k=0}^{\infty}\frac{z^{k}}
                                 {\Gamma(\alpha k+\beta)}, \quad
                                  \alpha, \beta \in \mathbb C,\, \operatorname{Re} (\alpha)>0,\, \operatorname{Re}(\beta)>0,\, z \in \mathbb C.
\end{align}
The main properties of the above mentioned functions, and other ML-type functions, can be found in the book by Erd\'{e}lyi \textit{et al.} \cite{Erdelyi1955}, 
and a detailed overview in the book by Dzhrbashyan \cite{Dzhrbashyan1966a}.
To demonstrate the concept of generality of the ML-type functions let us point out, that the ML function for one parameter \eqref{eq:1term-ML}, is a special case of the two-parameter ML function, \textit{i.e.} if we substitute $\beta = 1$ in \eqref{eq:2term-ML}. Accordingly, the classical exponential function is a~special case of the one-parameter ML function, where $\alpha = 1$:
\begin{eqnarray*}
     E_{\alpha,1}(z) =  \sum_{k=0}^{\infty}\frac{z^{k}}{\Gamma(\alpha k+1)}
                 \equiv  &E_{\alpha}(z),  \\ \nonumber
                  &  E_{1}(z) &=  \sum_{k=0}^{\infty}\frac{z^{k}}{\Gamma(k+1)}=
    \sum_{k=0}^{\infty}\frac{z^{k}}{k!}=e^{z} .
           \end{eqnarray*}

In 1971 the generalisation of the two-parameter function of the ML-type \eqref{eq:2term-ML} was introduced by T.R. Prabhakar \cite{Prabhakar1971} in terms of the series representation:
 \begin{align}\label{eq:3term-ML}
    E^{\gamma}_{\alpha, \beta}(z) = \sum_{k=0}^{\infty}\frac{(\gamma)_{k}z^{k}}
                                 {\Gamma(\alpha k+\beta)\, k!}, \quad
                                  \alpha, \beta, \gamma \in \mathbb C,\, \operatorname{Re} (\alpha)>0,\, \operatorname{Re} (\beta)>0,\, z \in \mathbb C,
\end{align}
where $(\gamma)_{k}$ is Pochhammer's symbol \cite{Rainville1960}, defined by:
 \begin{align*}
    (\gamma)_{k} = \frac{\Gamma(\gamma+k)}{\Gamma(\gamma)} = 
    \left\{ \begin{array}{ll} 1,& (k=0,\, \gamma \neq 0), \\
                                 \gamma(\gamma+1) \ldots (\gamma+k-1),& (k \in \mathbb N,\, \gamma \in \mathbb C).
                                 \end{array} \right.
        \end{align*}
The function defined in \eqref{eq:3term-ML} is a natural generalisation of the exponential function $e^z$, the one-parameter ML function $E_{\alpha}(z)$ and the Wiman's function $E_{\alpha, \beta}(z)$. 
In 2007 Shukla and Prajapati \cite{Shukla2007} proposed and investigated the function $ E^{\gamma,q}_{\alpha, \beta}(z)$, defined as:
 \begin{align*}
    E^{\gamma,q}_{\alpha, \beta}(z) = \sum_{k=0}^{\infty}\frac{(\gamma)_{qk}z^{k}}
                                 {\Gamma(\alpha k+\beta)\, k!}, \quad
                                  \alpha, \beta, \gamma \in \mathbb C,\, \operatorname{Re} (\alpha)>0,\, \operatorname{Re} (\beta)>0,\,\operatorname{Re} (\gamma)>0,\,  z \in \mathbb C,
\end{align*}
where $q \in  (0,1)\ \cup \ \mathbb N$, and $(\gamma)_{qk}$ denotes the generalised Pochhammer's symbol:
 \begin{align*}
    (\gamma)_{qk} = \frac{\Gamma(\gamma+qk)}{\Gamma(\gamma)},
        \end{align*}
which in particular reduces to:
 \begin{align*}
    q_{qk} = \prod_{r=1}^{q} \left(   
    \frac{\gamma+r-1}{q}\right)_{k}, \quad \text{if } q \in  \mathbb N.
        \end{align*}

Also other authors introduced and investigated further generalisations of the ML function, but to demonstrate the potential of the ML function these four generalisations of the exponential function are sufficient.

\section{Modelling the Phillips curve}
As in many fields of science and applications so in economics, to describe a relation between two variables the regression analysis is often used. One can use different regression models from simple linear-type, throughout exponential- and power-type models, to polynomial ones, and many other more complex and sophisticated. 
The discussion on the linearity or nonlinearity, and on the convex or concave shape of the PC, if it is supposed to be nonlinear, is still ongoing. Some authors are in favour of convex shape \cite{Cover1992, Karras1996, Nobay2000, Schaling2004}, 
some of concave \cite{Stiglitz1997}, and some of their combination \cite{Filardo1998}. The application of the ML-type function to describe the PC perfectly fits into this discussion.

\subsection{The ``Original'' Phillips Curve}
\begin{figure}[!htb]
\centering
\noindent
 	\includegraphics[width=0.75\textwidth]{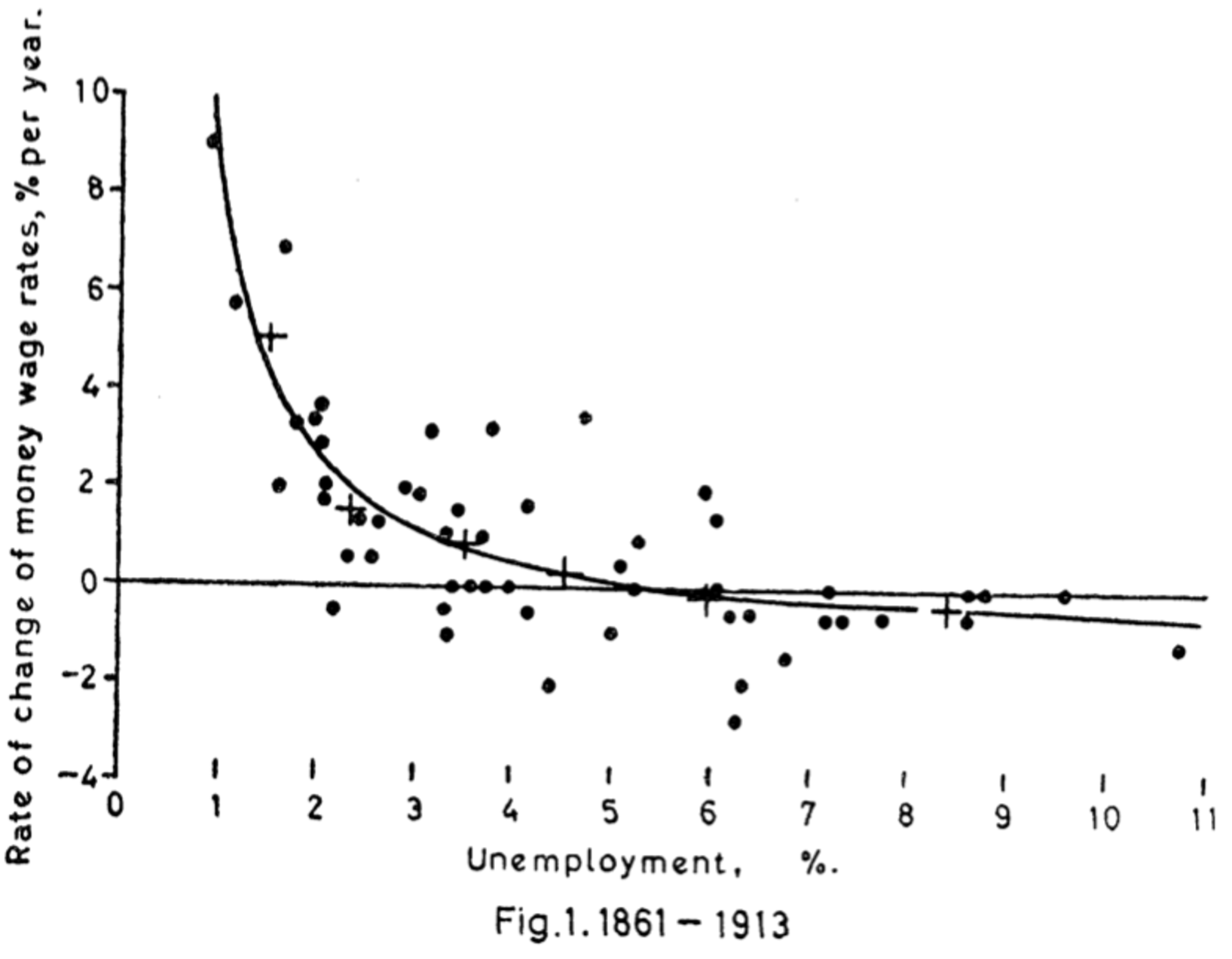}
   \caption{``Original'' Phillips curve \cite{Phillips1958}.}
   \label{fig:Phillips_1}
\end{figure}

Phillips in \cite{Phillips1958} used British econometric data - the rate of change of money wage rates, provided by the Board of Trade and the Ministry of Labour (calculated by Phelps Brown and Sheila Hopkins \cite{Brown1950}), and corresponding percentage employment data, quoted in Beveridge, Full Employment in a~Free Society, Table~22. 
But, for a simpler evaluation, the data were first preprocessed, \textit{i.e.} the average values of the rate of change of money wage rates and of the percentage unemployment for 6 different levels of the unemployment (0-2, 2-3, 3-4, 4-5, 5-7, 7-11) were calculated. The crosses in the Figure~\ref{fig:Phillips_1} refer to these average values. Each cross gives an approximation to the rate of change of wages which would be associated with the indicated level of unemployment if unemployment were held constant at that level. Finally, Phillips fitted a curve to the crosses using a model in the form:
 \begin{align}\label{eq:orig_PC_1}
    y + a  &=  b\, x^c \Rightarrow \\ \nonumber
    \log\, (y+a) &= \log\, b +c\, \log\, x,
     \end{align}
where $y$ is the rate of change of wage rates and $x$ the percentage unemployment. The constants $b$ and $c$ were estimated using the least squares to fit four crosses laying between 0-5 \% of unemployment, and constant $a$ was chosen to fit the remaining two crosses laying in the interval 5-11 \% of unemployment. Based on this ``fitting criterion'' Phillips identified the parameters of the model~\eqref{eq:orig_PC_1} as follows:
 \begin{align*}
    y + 0.900 & =  9.638\, x^{-1.394} \Rightarrow \\
    \log\, (y+0.900) &= 0.984 -1.394\, \log\ x.  
     \end{align*}

\subsection{The Mittag-Leffler Model for Fitting the Phillips Curve}
The idea to use a ML-type function to describe the econometric data (representing the Phillips curve) results naturally from observation of two facts:
\begin{itemize}
\item the simplicity of the model used by Phillips in his paper \cite{Phillips1958} given in \eqref{eq:orig_PC_1}, where a~power-type regression is used to fit the data, where the model can be defined in the form:
		 \begin{align}\label{eq:power_mod}
    			y(x)  =  b\, x^{c} - a, \quad a,b,c \in \mathbb R,
     		 \end{align}
\item the usual shape of the PC, used in the literature, which reminds on the exponential-type function:
		 \begin{align}\label{eq:exp_mod}
    			y(x)  =  b \operatorname{e}^{c\,x} - a, \quad a,b,c \in \mathbb R,
     		 \end{align}
\end{itemize}
where for both cases, \eqref{eq:power_mod} and \eqref{eq:exp_mod}, $x$ stands for the unemployment rate and $y$ for the inflation rate.

\begin{figure}[!htb]
\centering
\noindent
  \subfigure[$y(x) = c\,E_{\alpha}\, ( b\,x^{\alpha}\, )$]{
	\includegraphics[width=0.48\textwidth]{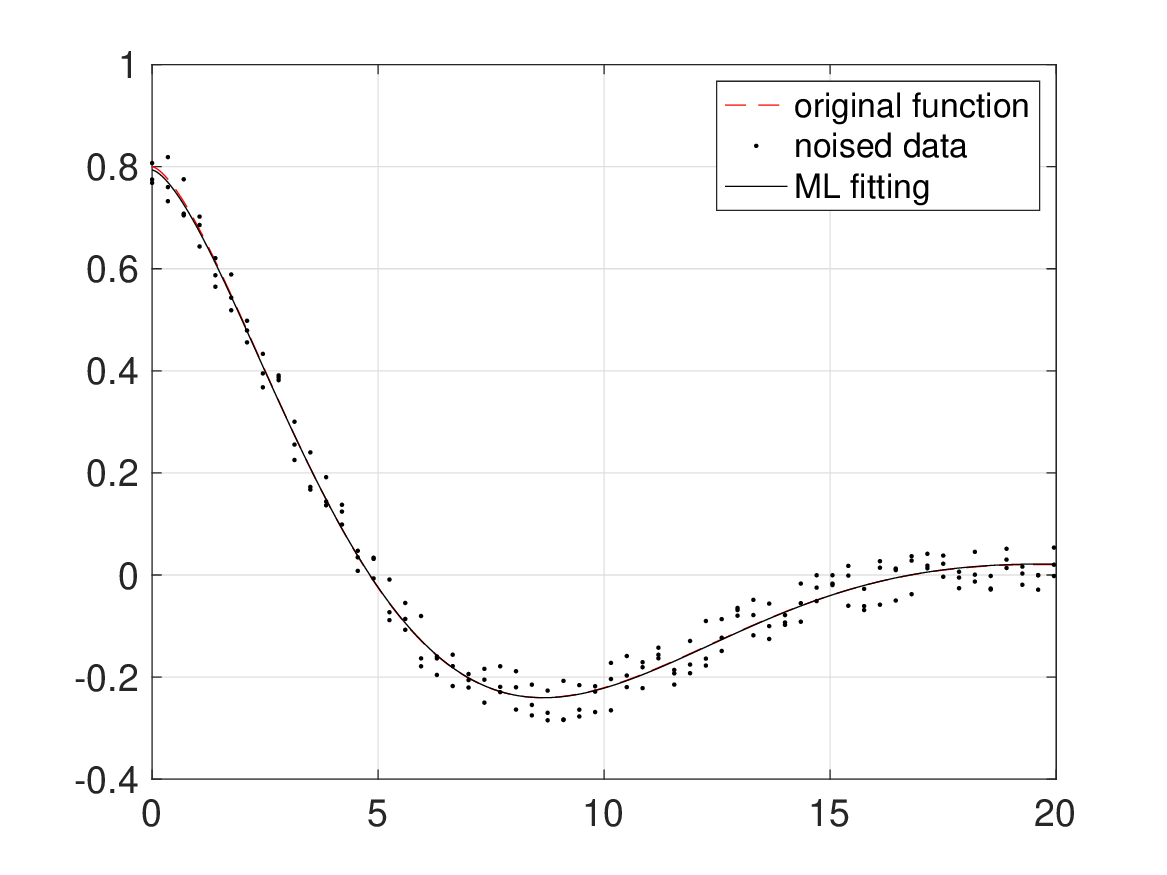}}
 \subfigure[$y(x) = e^{x}\, \operatorname{erfc}\,(\sqrt{x})$]{
               \includegraphics[width=0.48\textwidth]{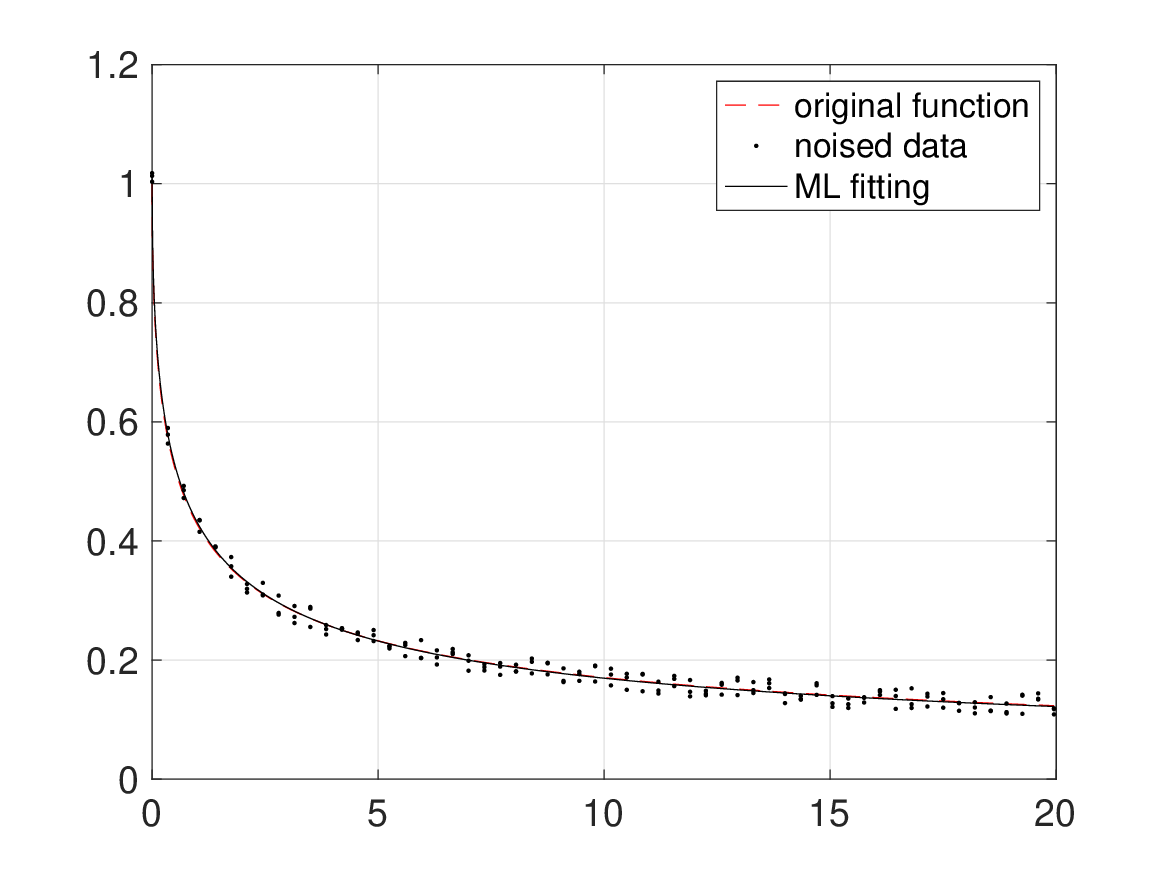}}
  \subfigure[ $y(x) = e^{- \alpha x} \cos\, (x)$]{
	\includegraphics[width=0.48\textwidth]{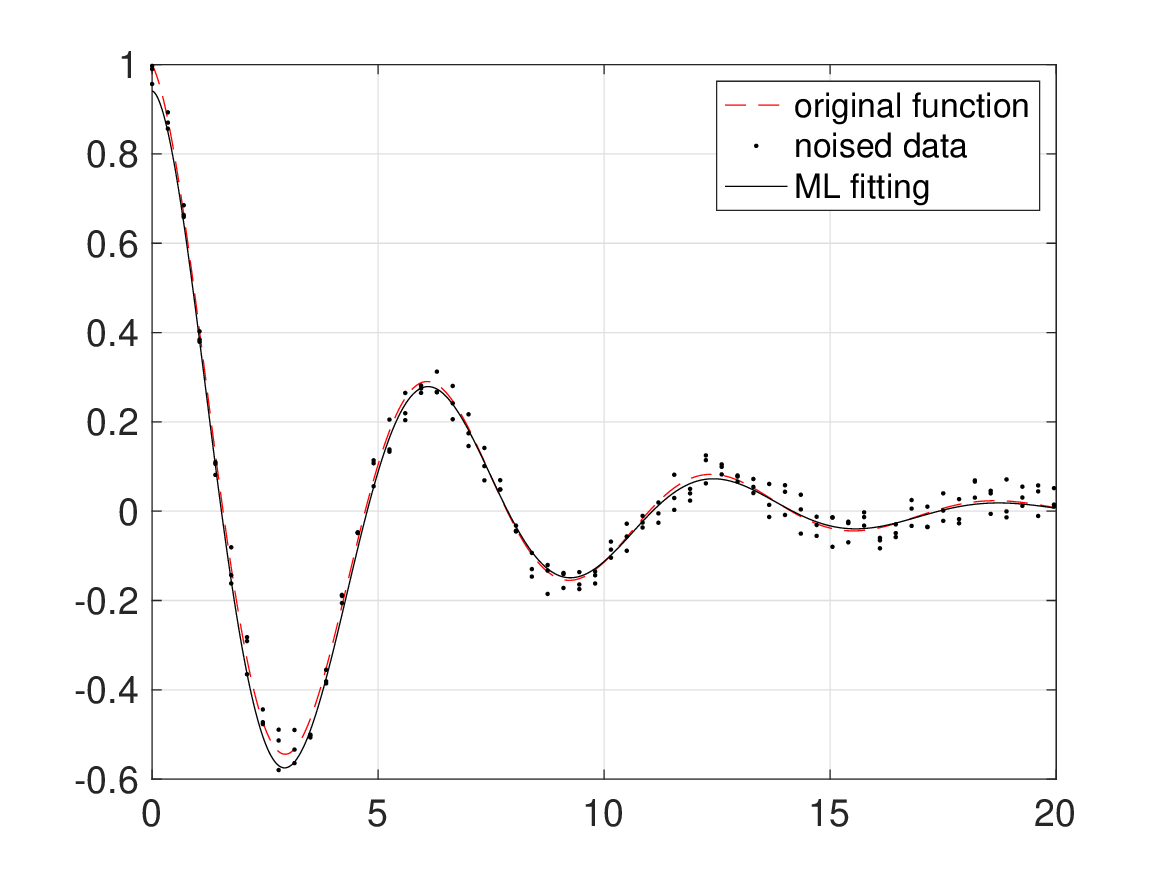}}
 \subfigure[$y(x) = \cos \, (x)$]{ 
              \includegraphics[width=0.48\textwidth]{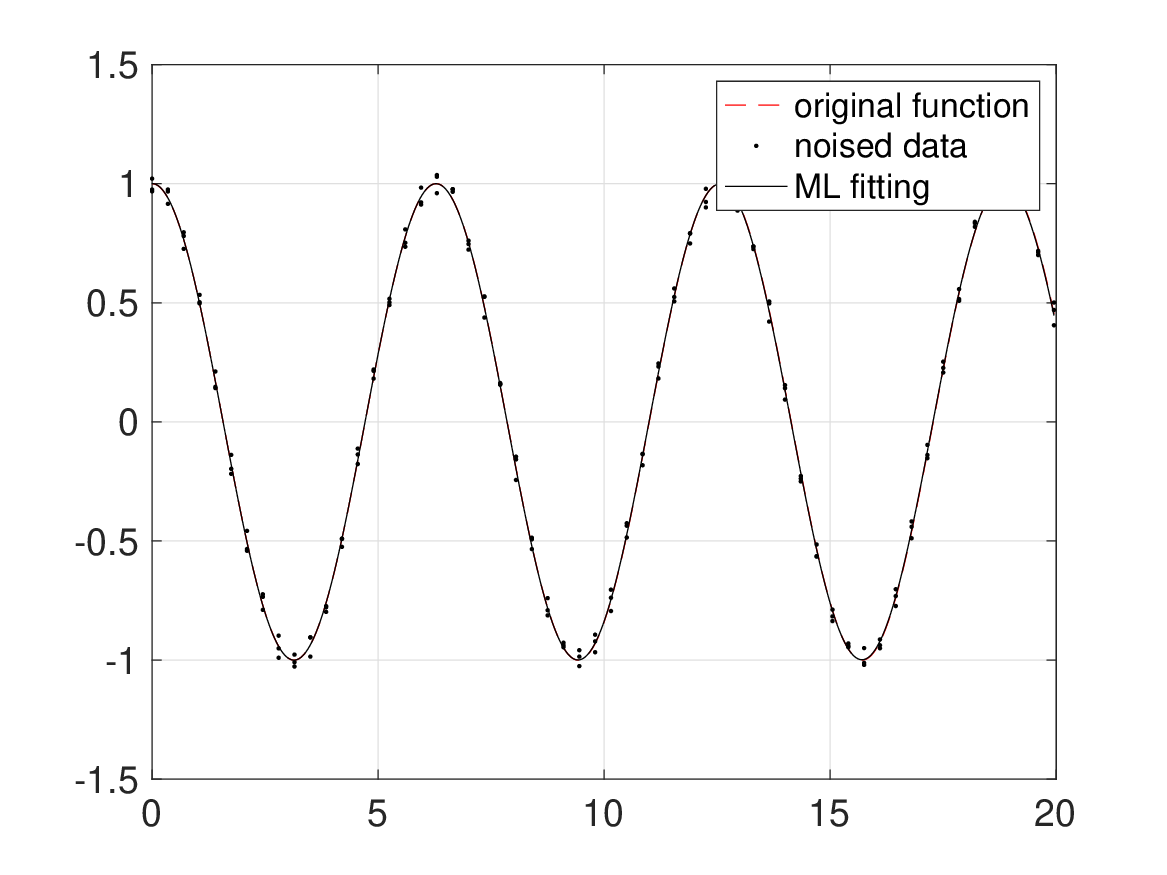}}
  \caption{Mittag-Leffler fitting using different functions $y(x)$ for generating data \cite{PodlubnyMLfit}.}
  \label{fig:ML_fit_1_2_3_4}
\end{figure}

Based on these facts, the one-parameter ML function appears to be a general model to fit the PC relation, as it behaves between a~purely exponential function and a~power function. Some of the possible manifestations of the ML function are shown in Figure~\ref{fig:ML_fit_1_2_3_4} (figures generated using the Matlab demo published by Igor Podlubny~\cite{PodlubnyMLfit}).

The one-parameter ML function defined in \eqref{eq:1term-ML}, which includes the special case when $\alpha =1$, \textit{i.e.} the classical exponential function, is used to model the econometric data under study. Generally, the proposed fitting model can be written as follows:
\begin{align}\label{eq:ML_model}
y(x) = c\, E_{\alpha}\, ( b\,x^{\alpha}\, ),\quad \alpha \in \mathbb C,\quad \operatorname{Re} (\alpha)>0,\quad b, c \in \mathbb R,
\end{align} 
where the parameters $\alpha, b, c$ are subject to optimisation procedure minimising the squared sum of the vertical offsets between the data points and the fitting curve. 
For the evaluation of the one-parameter ML function, the Matlab functions created by Podlubny, Chen and Garrappa \cite{PodlubnyML, PodlubnyMLfit, ChenML,ChenML2, GarrappaMAT} were used, all giving identical results.

\section{Numerical Results and Discussion}
\label{sec:fitting_cases}

To evaluate the performance of the proposed ML model \eqref{eq:ML_model} in comparison to the power-type model \eqref{eq:power_mod}, and the exponential-type model \eqref{eq:exp_mod} the econometric data of two European countries (France and Switzerland) were used, that were obtained from the \emph{EconStats\textsuperscript{TM}} portal \cite{econstats}. The unemployment rate and inflation rate were taken for the period of time 1980-2017. The whole list of the  processed data can be found in the Table~\ref{Tab:original}. 

\subsection{Goodness-of-Fit Statistics and Data Preprocessing}
The sum of squared errors (SSE) between the fitting models and the used data serves as the fitting-criterion, with values closer to $0$ indicating a smaller random error component of the model. Also some other quality measures were evaluated, \textit{i.e.} the R-square from interval $[0,\ 1]$, with values closer to $1$ indicating that a greater proportion of variance is accounted for by the model (\textit{e.g.} value of $0.7325$ means that the fit explains $73.25\%$ of the total variation in the data about the average); the adjusted R-square statistic, with values smaller or equal to $1$, where values closer to $1$ indicate a better fit; the root mean squared error (RMSE), with values closer to $0$ indicating a fit that is more useful for prediction \cite{MathWorks}.

The used dataset, where the unemployment rate corresponds to the $x$-coordinate and the inflation rate corresponds to the $y$-coordinate of each sample representing the state of these two indicators for each year from the period under study, is first split into two subsets, the ``modelling'' subset is used to identify the model parameters, the ``out-of-sample'' subset serves for evaluating the forecast-performance of the models.
For both economies, French ad Swiss, all three models were first fitted to the data from the ``modelling'' subset (composed of 31 samples), by minimising SSE, identifying the optimal parameters. The obtained parameters were then used to compute SSE of the identified models to the ``out-of-sample'' subset  (composed of 7 samples with the greatest values of unemployment rate) and SSE of the fitting model to the complete dataset.

\subsection{Experiments}

The first experiment was conducted using the French econometric data. The ``modelling'' subset of 31 samples, was used for the identification purposes. All three models, the power-type model \eqref{eq:power_mod}, the exponential-type model \eqref{eq:exp_mod}, and  the ML model \eqref{eq:ML_model}, were fitted to these data minimising the SSE obtaining so the optimal parameters. 
The identified models were then used to compute the SSE to the complete dataset of 38 samples (including the ``out-of-sample'' subset).
SSE results to the ``modelling'' subset as well as SSE to the ``out-of-sample'' subset and  SSE to the complete dataset for the French Phillips Curve are shown in Table~\ref{tab:FR}, alongside the values of R-square, adjusted R-square, and RMSE.
 The ML model outperforms the compared models in all listed statistic indicators, with SSE to the ``out-of-sample'' subset double smaller than the exponential-type model, and almost three-times smaller than the power-type model (see Table~\ref{tab:FR}, where bold stands for better result). 
 \begin{table}[!h]
\caption{The statistical results of the French Phillips curve fitting.}
\centering
\small
\label{tab:FR}
\begin{tabular}{l|ccc}
\hline
\phantom{} &	power-type model	&	exponential-type model	&	ML model	\\
\hline
SSE to ``modelling’’ subset	&	157.8422	&	155.8276	&	\bf{149.6035}	\\
SSE to ``out-of-sample’’ subset	&	10.9024	&	    8.0347	&	\bf{3.9904}	\\
SSE to complete dataset	&	168.7446	&	163.8623	&	\bf{153.5939}	\\
R-square	&	0.5634	&	0.569	&	\bf{0.5862}	\\
adjusted R-square	&	0.5322	&	0.5382	&	\bf{0.5567}	\\
RMSE	&	2.374	&	2.359	&	\bf{2.311}	\\
\hline
Model		 &	$y(x)  =  b\, x^{c} - a$  &	$y(x)  =  b \operatorname{e}^{c\,x} - a$ &	$y(x) = c\, E_{\alpha}\, ( b\,x^{\alpha}\, ) $ \\
definition &&&\\
				& $a= 1.552 $ & $a = -0.0187$ & $\alpha = 1.358$  \\ 
Identified  		 	& $b = 1.009e+04$ & $b = 563.6$ & $b = -0.6378$ \\ 
parameters		& $c = -3.471$ & $c = -0.571$ & $c = -149.9$  \\
		   \hline
		   \end{tabular}
\end{table}

\begin{figure}[!htb]
\centering
	\includegraphics[width=0.75\textwidth]{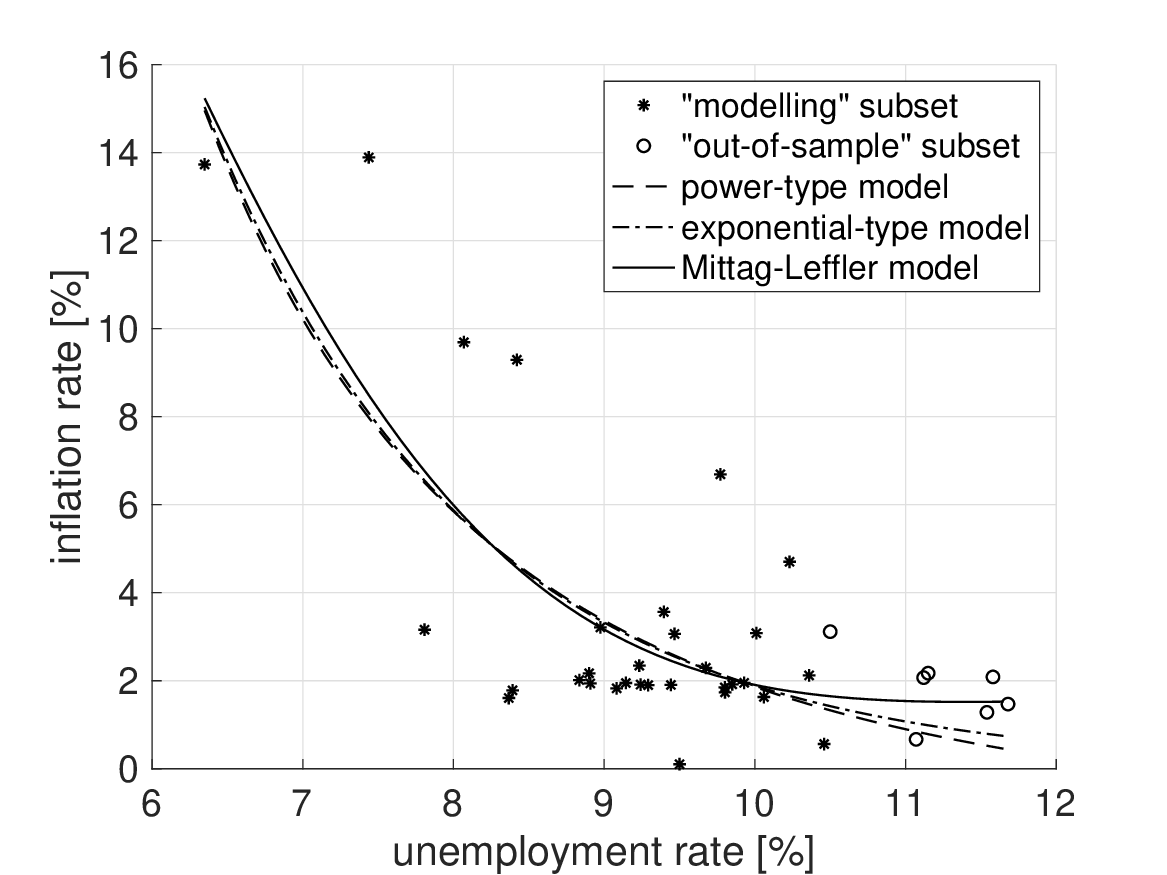}
   \caption{Fitting the French Phillips curve.}
   \label{fig:FR}
\end{figure}

The result of the French Phillips Curve fitting is also shown in Figure~\ref{fig:FR}, where it is possible to observe a similar behaviour of all three models, \textit{i.e.} with the increase of the unemployment rate the inflation rate exponentially decreases. However, for the last samples, the decrease of the ML model slows down in comparison to the power-type and exponential-type models. This behaviour of the ML model is obviously better copying the trend of the ``out-of-sample'' subset. This is also confirmed by the smallest SSE value of the ML fitting curve to the ``out-of-sample'' subset (see Table~\ref{tab:FR}).

Identically as in the French case, the Swiss econometric data (unemployment rate and inflation rate) were first preprocessed. The complete dataset was split into the ``modelling'' subset composed of 31 samples, that was used to identify the optimal parameters of the power-type model \eqref{eq:power_mod}, the exponential-type model \eqref{eq:exp_mod}, and  the ML model \eqref{eq:ML_model}. The SSE between the ``modelling'' subset and the fitting curves was again used as the fitting criterion. Using the identified model parameters the SSE of the evaluated models to the complete dataset of 38 samples (including the ``out-of-sample'' subset) was computed. In order to compare the forecast-performance of the models
SSE  to the ``out-of-sample'' subset, as well as the SSE values for the ``modelling'' subset fitting, and SSE to the complete dataset for the Swiss Phillips Curve are shown in Table~\ref{tab:SUI}, alongside the values of R-square, adjusted R-square, and RMSE.
\begin{table}[!h]
\caption{The statistical results of the Swiss Phillips curve fitting.}
\centering
\small
\label{tab:SUI}
\begin{tabular}{l|ccc}
\hline
\phantom{} &	power-type model	&	exponential-type model	&	ML model	\\
\hline
SSE to ``modelling’’ subset	&	39.6506	&	40.0588	&	\bf{39.2992}	\\
SSE to ``out-of-sample’’ subset	&	6.8961	&	\bf{4.6826}	&	5.0041	\\
SSE to complete dataset	&	46.5466	&	44.7414	&	\bf{44.3033}	\\
R-square	&	0.6389	&	0.6351	&	\bf{0.642}	\\
adjusted R-square	&	0.6131	&	0.6091	&	\bf{0.6165}	\\
RMSE	&	1.19	&	1.196	&	\bf{1.185}	\\
\hline
Model		 &	$y(x)  =  b\, x^{c} - a$  &	$y(x)  =  b \operatorname{e}^{c\,x} - a$ &	$y(x) = c\, E_{\alpha}\, ( b\,x^{\alpha}\, ) $ \\
definition &&&\\
				& $a= -551.3 $ & $a = -0.7809$ & $\alpha = 0.7733$  \\ 
Identified  		 	& $b = -548.6$ & $b = 6.364$ & $b = -1.468$ \\ 
parameters		& $c = 30.85e-04$ & $c = -1.376$ & $c = 8.823$  \\
		   \hline
\end{tabular}
\end{table}

\begin{figure}[!htb]
\centering
                \includegraphics[width=0.75\textwidth]{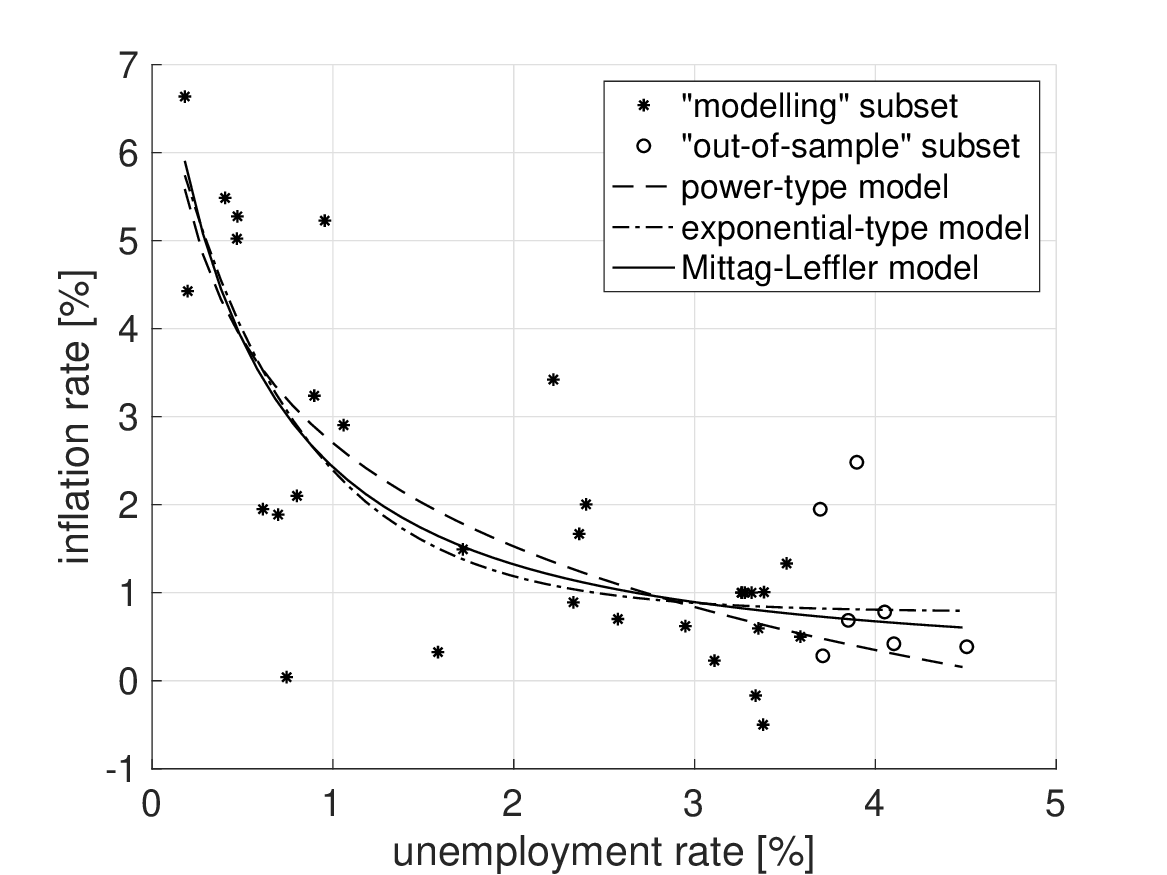}
   \caption{Fitting the Swiss Phillips curve.}
   \label{fig:SUI}
\end{figure}

Observing the result of the Swiss Phillips Curve fitting shown in Figure~\ref{fig:SUI}, one can see an interesting case, where although all the compared models are exponentially decreasing, the curve representing the proposed ML model proceeds in-between the power-type and the exponential-type models, that form a kind of scissors. In respect to the ``out-of-sample'' subset it is possible to observe that two points of that subset deviate, having higher inflation rate value then the others. This strongly influenced the fitting results. In this case the exponential-type model visually represents the ``out-of-sample'' subset slightly better then the ML model, that is also demonstrated by a smaller value of SSE of the exponential-type model to the ``out-of-sample'' subset (see Table~\ref{tab:SUI}). In spite of this, the ML model outperforms the compared models in all other used statistic indicators, including smaller SSE to the complete dataset, proving it's capability. Moreover, in case of filtering these two outliers from the ``out-of-sample'' subset, the ML model better fits the data-trend.

\section{Conclusion}
\label{sec:Concl}

The ability of the Mittag-Leffler function to behave between the power-type and the exponential-type function, and moreover to fit data that manifest signs of stretched exponentials, oscillations or damped oscillations is demonstrated in this paper, with application to fitting the econometric data (Phillips curve) of two European economies. Exploiting the full potential of the Mittag-Leffler function and it's generalisations, as well as associating the model parameters with the corresponding economic indicators will be the topic of further work.

\vspace{6pt} 

\section*{Acknowledgements}
{This research was funded in part by the Slovak Research and Development Agency under
Grants APVV-14-0892, SK-SRB-18-0011, SK-AT-2017-0015, APVV-18-0526;
 in part by
the Slovak Grant Agency for Science under Grant VEGA 1/0365/19; and
in part by the framework of the COST Action CA15225.
}

\section*{}
The author declare no conflict of interest.


\newpage
\textbf{Appendix: The econometric dataset}

\begin{table}[!ht]
\small
\caption{The complete dataset: Econometric data for years 1980 -- 2017 \cite{econstats}.}
\label{Tab:original}
\centering
\label{tab:original_data}
\begin{tabular}{c|cc|cc}
\hline\phantom{1980}  & \multicolumn{2}{|c|}{\textbf{France}} & \multicolumn{2}{|c}{\textbf{Switzerland}} \\
Year &	unemployment	&	inflation  &	unemployment	&	inflation \\
 & rate [\%]& rate [\%] & rate [\%]& rate [\%] \\
\hline
1980	&	6.3490	&	13.7300	&	0.1970	&	4.4260	\\ 
1981	&	    7.4380	&	   13.8900	&	    0.1810	&	    6.6370	\\ 
1982	&	    8.0690	&	    9.6910	&	    0.4040	&	    5.4850	\\ 
1983	&	    8.4210	&	    9.2920	&	    0.8010	&	    2.1000	\\ 
1984	&	    9.7710	&	    6.6900	&	    1.0590	&	    2.9040	\\ 
1985	&	   10.2300	&	    4.7030	&	    0.8970	&	    3.2380	\\ 
1986	&	   10.3600	&	    2.1210	&	    0.7440	&	    0.0400	\\ 
1987	&	   10.5000	&	    3.1150	&	    0.6970	&	    1.8870	\\ 
1988	&	   10.0100	&	    3.0810	&	    0.6130	&	    1.9490	\\ 
1989	&	    9.3960	&	    3.5630	&	    0.4690	&	    5.0220	\\ 
1990	&	    8.9750	&	    3.2120	&	    0.4720	&	    5.2760	\\ 
1991	&	    9.4670	&	    3.0630	&	    0.9550	&	    5.2270	\\ 
1992	&	    9.8500	&	    1.9180	&	    2.2190	&	    3.4210	\\ 
1993	&	   11.1200	&	    2.0700	&	    3.8970	&	    2.4820	\\ 
1994	&	   11.6800	&	    1.4690	&	    4.1020	&	    0.4200	\\ 
1995	&	   11.1500	&	    2.1720	&	    3.6950	&	    1.9480	\\ 
1996	&	   11.5800	&	    2.0860	&	    4.0510	&	    0.7810	\\ 
1997	&	   11.5400	&	    1.2820	&	    4.5050	&	    0.3860	\\ 
1998	&	   11.0700	&	    0.6680	&	    3.3380	&	   -0.1680	\\ 
1999	&	   10.4600	&	    0.5620	&	    2.3620	&	    1.6680	\\ 
2000	&	    9.0830	&	    1.8270	&	    1.7190	&	    1.4930	\\ 
2001	&	    8.3920	&	    1.7810	&	    1.5810	&	    0.3250	\\ 
2002	&	    8.9080	&	    1.9380	&	    2.3300	&	    0.8910	\\ 
2003	&	    8.9000	&	    2.1690	&	    3.3530	&	    0.5940	\\ 
2004	&	    9.2330	&	    2.3420	&	    3.5090	&	    1.3320	\\ 
2005	&	    9.2920	&	    1.9000	&	    3.3840	&	    1.0060	\\ 
2006	&	    9.2420	&	    1.9120	&	    2.9490	&	    0.6210	\\ 
2007	&	    8.3670	&	    1.6070	&	    2.4000	&	    2.0040	\\ 
2008	&	    7.8080	&	    3.1590	&	    2.5760	&	    0.7010	\\ 
2009	&	    9.5000	&	    0.1030	&	    3.7090	&	    0.2830	\\ 
2010	&	    9.8020	&	    1.7360	&	    3.8500	&	    0.6860	\\ 
2011	&	    9.6750	&	    2.2930	&	    3.1100	&	    0.2280	\\ 
2012	&	    9.9290	&	    1.9520	&	    3.3790	&	   -0.5000	\\ 
2013	&	   10.0600	&	    1.6300	&	    3.5850	&	    0.5000	\\ 
2014	&	    9.8010	&	    1.8480	&	    3.3150	&	    1.0000	\\ 
2015	&	    9.4430	&	    1.9040	&	    3.2780	&	    1.0000	\\ 
2016	&	    9.1440	&	    1.9490	&	    3.2590	&	    1.0000	\\ 
2017	&	    8.8350	&	    2.0150	&	    3.2620	&	    1.0000	\\ 
\hline
\end{tabular}
\end{table}

\end{document}